\documentclass{elsart5p}
\usepackage{graphicx, amsmath, amssymb, natbib}
\begin{document}
\begin{frontmatter}
%
\title{The information capacity of hypercycles}
%
\author[IFSC]{Daniel  A. M. M. Silvestre,}
\author[IFSC]{Jos\'e F. Fontanari\corauthref{CA}}
\corauth[CA]{Tel.: +55 1633739849; fax: +55 1633739877; fontanari@ifsc.usp.br}
\address[IFSC]{Instituto de F\'{\i}sica de S\~ao Carlos, Universidade de S\~ao Paulo, Caixa Postal 369, 
13560-970 S\~ao Carlos SP, Brazil}
%

%
\begin{abstract}

Hypercycles are information integration systems  which are  thought to overcome the information crisis of prebiotic evolution
by ensuring the coexistence of several short templates.  For imperfect template replication, we derive a simple 
expression for the maximum number of distinct templates $n_m$ that can coexist
in a hypercycle and show that it is a decreasing function
of the length $L$ of the templates. In the case of high replication accuracy 
we find that the product $n_m L$ tends to a constant value, limiting thus the information content of the hypercycle.
Template coexistence is  achieved either as a stationary equilibrium (stable fixed point) or
a stable periodic orbit in which the total concentration of functional templates is nonzero.
For the hypercycle system studied here we find numerical evidence  that  the existence of an unstable fixed point
is a necessary condition for the presence of periodic orbits. 

\end{abstract}
%
\end{frontmatter}

\section{Introduction}\label{sec:1}

Most of the modern theoretical work on prebiotic evolution was prompted by the seminal paper
of \citet{Eigen1971} which explored the fate of a population of competing macromolecules in
an environment with limited resources. The main conclusion of Eigen's work is that the
length of a replicating polymer (i.e., a RNA-like template) is limited by the 
replication accuracy per nucleotide, and so primordial replicators would have
to replicate with implausible high accuracy in order to reach the length of today's
RNA viruses (about $10^{3}$ to $10^{4}$ nucleotides).  This finding, together
with the observation that distinct templates cannot coexist 
in the competition-only scenario \citep{Swetina1982}, come to be known as the 
information crisis of prebiotic evolution.  

In fact, this information crisis would be resolved by the coexistence of several distinct,
short templates, i.e. by the splitting of  the information in short modules, similarly
to the   division of the genome in chromosomes found in many present-day organisms. In this case, 
the total information content of 
the template pool is the product of the number of different templates and the maximum information 
coded per template (roughly the template length $L$), 
provided the template types have the same concentration. But template coexistence can be achieved
only if some sort of  cooperation between templates is imposed \textit{a priori} to the
molecular population.
To this end \citet{Eigen1979}  proposed a
cyclic reaction scheme, termed hypercycle, in which each replicator would aid in the 
replication of the next one, in a regulatory cycle closing on itself. An alternative
proposal confines the templates in
packages  or prebiotic vesicles which are deemed viable provided it encloses a number $n$ of
distinct functional templates \citep{Niesert1981,Szathmary1987,Zintzaras2002,Fontanari2006}. 

Perhaps because hypercycles and vesicle models are  more
difficult to analyze than the naked-gene (quasispecies) scenario, the all-important question
of whether these information integration systems exhibit a similar phenomenon as the error threshold 
of the quasispecies model was put off. Only recently a re-examination of a prototypical package model
-- the model of \citet{Niesert1981} -- revealed that package models in general
suffer from the same malady as the quasispecies model: in the case of imperfect replication
an increase in the number $n$ of distinct
templates confined in the vesicle must be followed by a decrease of their lengths,
otherwise the package
becomes unviable. As a result, the product $nL$ (i.e., 
the total information content of the package) 
tends to a constant value that depends essentially on the spontaneous mutation rate per 
nucleotide \citep{Silvestre2007,Silvestre2008}. Our aim in this contribution is to investigate whether 
a similar restriction to the total amount of information in the pool of templates holds for 
hypercycle systems as well.

The dynamics of hypercycle systems in the presence of a variety of mutant types was extensively investigated by 
\citet{Stadler1992,Happel1998}. We refer the reader to \citet{Bresch1980,Szathmary1987,Sardanyes2007} for 
an emphasis on the destabilizing effects of mutant parasites and to \citet{Boerlijst1991,Cronhjort1994} for
an analysis of the robustness  conferred by spatial organization against those mutants.
However, the formulation of \citet{Campos2000} in which the 
mutants form an error tail class
seems more appropriate to study the error threshold phenomenon and so our analysis  will build heavily on 
that paper. 

The sole motivation of this contribution is to show that the condition
for the viability of the hypercycle  derived in \citet{Campos2000} is in fact valid for all $n$
and not only for the regime where fixed points are stable, i.e., for $n \leq 4$. This is so because we
found numerically that a necessary condition for the presence  of stable periodic orbits is the
existence of an unstable fixed point.  For the purpose of completeness, 
in the following section we describe the model and re-derive the main results regarding the
existence of fixed point (and hence of stable periodic solutions) for the hypercycle system.

\section{Model}

The hypercycle system we consider here is  composed of $n$ `functional' elements $I_1,
\ldots, I_n$ and its error tail $I_e$.   The templates are capable of self-replication
with productivity values $A_i ~\left( i=1, \ldots, n \right )$ and
$A_e$.  As usual, we introduce  the kinetic
constants $K_i$ that measure the strength of the influence of template $I_{i -1}$ on the growth promotion 
of template $I_i$. 
The key ingredient in the modeling is that in both
processes of growth of template $I_i$ the probability of success is given
by the parameter $Q \in [0,1]$, so that an erroneous copy, which will then
belong to the error tail, is produced with probability $1-Q$. Back-mutations from the error
class to the functional class, as well as  mutations between elements of the functional class,
are neglected. This formulation is equivalent to considering polynucleotides
of length $L \rightarrow \infty$ whose mutation probability per nucleotide $u$ 
 goes to $0$ such that the replication accuracy per genome
is finite, i.e.  $\exp \left ( -Lu \right ) \rightarrow Q$.

The
concentrations $x_i~\left ( i=1, \ldots,n \right )$ of the hypercycle
functional elements and the concentration $x_e$ of the error-tail evolve in time
according to the kinetic equations \citep{Campos2000}
\begin{equation}\label{ode_h}
\dot x_{i} = x_{i} \left ( A_{i}Q + K_i x_{i-1}Q - \Phi \right ) 
~~~~~~~ i=1,...,n
\end{equation}
and
\begin{equation}\label{ode_e}
\dot x_{e} = x_{e} \left ( A_{e}  - \Phi \right ) + \left ( 1-Q \right ) 
\sum_{i=1}^{n} x_{i} \left ( A_{i}  + K_i x_{i-1} \right )  
\end{equation}
where $x_0 \equiv x_n$ and
\begin{equation}\label{flux}
\Phi= \sum_{i=1}^n x_i \left ( A_i + K_i x_{i-1} \right ) + x_e A_e
\end{equation}
is a dilution flux that keeps the total concentration constant, i.e.,
$ \sum_{i=1}^{n} \dot x_{i} + \dot x_{e}  = 0$. As usual,
the dot denotes a time derivative. Henceforth we will assume
that 
\begin{equation}\label{CC}
 \sum_{i=1}^{n} x_{i} +x_{e} = 1 .
\end{equation}
In accord with the usual assumption of 
package models that functional templates and parasites are
selectively neutral \citep{Niesert1981,Silvestre2008}
we  set $A_i = A_e = a$, and  $K_i=K> 0$ for
$i=1, \ldots,n$, resulting in the so-called symmetric hypercycle. Hence by
measuring the kinetic constant $K$ and the time $t$ in units of $a$ we can set
$a=1$ without loss of generality (except for $a=0$, of course).

\section{Analysis of the steady state}

Here we focus only on the fixed-point solutions $x_i > 0$ for $i=1,\ldots,n$. In this case,
the condition $\dot x_2 = 0$ yields $\Phi = Q + K Q x_1$ which, inserted in the
equations $\dot x_3 = \ldots = \dot x_n = \dot x_1 =0$, yields $x_1 = x_2 = \ldots
= x_{n}$. Using these results in Eq. (\ref{flux}) we find that $x_1$
is given by the roots of the quadratic equation
\begin{equation}\label{quad}
n  K  x_1^2 - K Q  x_1 + 1 - Q = 0, 
\end{equation}
which has two real positive roots provided the condition
\begin{equation}\label{Qh}
 K Q^2 - 4n \left (1 -Q\right )  \geq 0
\end{equation}
is satisfied. We note that $K>0$ is a necessary condition for the existence
of hypercyclic solutions $x_1 = \ldots = x_n > 0$, since in the absence of catalytic couplings 
among  functional templates (i.e., $K=0$) Eq. (\ref{quad})
has no solution.

The analysis of the roots of Eq.\ (\ref{quad})
and the numerical evaluation of the Jacobian eigenvalues indicate that
the smaller root is always unstable whereas the larger root is
(locally) stable for $n \leq 4$. In addition, the disordered fixed point
$x_i = 0, \forall i$ and $x_e = 1$ is always stable \citep{Campos2000}.

\begin{figure}
\centerline{
\resizebox{0.75\columnwidth}{!}{\includegraphics{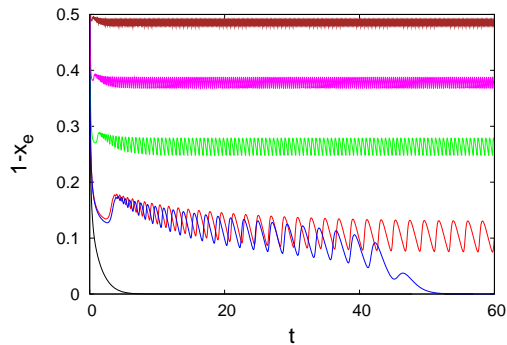}}}
\caption{Time evolution of the total concentration of functional templates for
$n=12$,  $K = 1000$ and (top to bottom) $Q= 0.5, 0.4, 0.3, 0.2, 0.19639, 0.1$.  The 
periodic solutions disappear at $Q  \approx 0.19639$, the value at which the
condition (\ref{Qh}) for the existence of a nontrivial equilibrium is violated. 
We have no proof for this remarkable coincidence. For lower
values of replication accuracy the dynamics converges to the trivial
fixed point $x_e=1$, which is always stable.
}
\label{fig:1}
\end{figure}

It is well known
that for $n >4$ the steady state of the hypercycle  is characterized by stable periodic
solutions \citep{Eigen1979,Hofbauer1988,Hofbauer1991} where the concentrations $x_i$ vary wildly reaching 
values dangerously close to zero, which would certainly doom a finite-population  
system \citep{Nuno1994}. Nevertheless, here we take an optimistic stance and assume
that the hypercycle is an acceptable information integrator even in the regime where it exhibits 
periodic solutions:  the $n$ templates do coexist in this regime after all, albeit
with (very) distinct concentrations. The issue is then to determine the region in the space of the parameters
$Q$ and $K$ where the steady-state solution is such that $\sum_{i=1}^n x_n = 1 - x_e$ is nonzero.

The search for steady-state solutions with  nonvanishing concentration of functional templates can be
carried out straightforwardly through the numerical integration of the differential equations  (\ref{ode_h})-
(\ref{ode_e}). The procedure is illustrated in Fig.\ \ref{fig:1} for $n=12$: for a fixed value of $K$ we
decrease $Q$ until we reach a threshold value below which the only steady-state solution is $x_e = 1$.

Remarkably, through this numerical analysis we find 
that the region of viability
of the hypercycle (i.e., the region characterized by a nonzero concentration of functional templates, regardless
of whether the stable solution is a fixed point or a periodic orbit) is determined by the condition of existence
of real fixed points, Eq. (\ref{Qh}). We are not aware of a mathematical proof for this result. For $n=2$, index theory
can be used to show that inside the region enclosed by a periodic orbit there must exist at least
one fixed point  so that the absence of fixed points precludes the existence of periodic orbits 
\citep[pp. 51]{Guckenheimer1983}. 
It seems, however, that this result cannot be extended to $n > 2$. The theorem proved by \citet{Hofbauer1991} is not helpful either: 
it asserts that for a hypercycle system with no error tail (i.e., $Q=1$)
 if the (unique) fixed point becomes unstable then there is a stable periodic orbit, 
whereas our conjecture is that if there is a stable periodic orbit then there must be at least one unstable fixed point.

In summary, we find that  the hypercycle system is viable provided
the number of functional templates $n$ satisfies the condition $n \leq n_m$ where
\begin{equation}\label{n_max}
n_m = \frac{K Q^2}{ 4 \left (1 -Q\right )} .
\end{equation}
Since $n_m$ is a monotonously increasing function of $Q \in [0,1]$ and 
$Q$, in turn, is a monotonously decreasing function of $L$ (recall that
$Q = \exp \left ( -uL \right )$) we find that $n_m$ decreases with increasing $L$.
Hence, all other things being equal if the number of functional templates $n$ is increased
then their lengths $L$ must decrease accordingly so as to guarantee that $n \leq n_m$ is
fulfilled. This remark shows that an information preservation principle similar to that
derived for package models holds for the hypercycle as well, which suggests a reconsideration of
the whole approach based on the coexistence
of distinct templates to address the information crisis of prebiotic evolution.

\section{Conclusion}

In our analysis we have opted for the choice of parameters that most favored  the
stability of the hypercycle. For instance, introduction of other interactions such as the catalytic promotion of 
the growth of the templates in the error tail by functional templates -- an
assumption implicit in the package models -- can only reduce the value of $n_m$. (We have verified that
the catalytic coupling between functional templates and the error tail does not produce any qualitatively 
new result.) Similarly, the parameter setting that corresponds to the
elementary hypercycle in which $A_i=0, \forall i$ but $A_e=1$ results also in a reduction of $n_m$.
Asymmetric hypercycles in which the templates have distinct productivities $A_i$ leads to internal
competition and again to the decrease of $n_m$ \citep{Campos2000}.
Hence Eq.\ (\ref{n_max}) must be seen as an upper bound to the maximum number of 
functional templates in the hypercycle. 

The effect that a change $\delta L$ causes on $n_m$ is given by
\begin{equation}\label{del}
\frac{\delta n_m}{n_m} = - u \left ( 2 + \frac{Q}{1-Q} \right ) \delta L .
\end{equation}
In the case the functional templates have a high replication accuracy
(i.e., $1 - Q \approx uL \approx 0$) we find that the product $n_m L$ tends to
a constant value, similarly to the findings for the package models \citep{Silvestre2007,Silvestre2008}.

We stress that Eq.\ (\ref{n_max}) is worthful because its validity 
is not restricted to the regime where the nontrivial fixed point is stable: it  holds also in the regime where the only stable
solutions are periodic orbits. We provide only  numerical evidence to support this remarkable finding
which is based on the conjecture that the existence of an unstable fixed point is a necessary condition for
the presence of stable periodic orbits in the system of ordinary differential  equations (\ref{ode_h})-(\ref{ode_e}).
It would be very interesting to find a proof for this conjecture.

\section{Acknowledgements}
D.A.M.M.S. is supported by CAPES. The work of J.F.F was supported in part by CNPq and FAPESP, Project No. 04/06156-3.

\end{document}